\documentclass[prl,showkeys,floatfix,twocolumn,amsmath,amssymb,floatfix]{revtex4}
\usepackage{graphicx,color,dcolumn,booktabs,bm}
\usepackage{longtable,lscape}
\usepackage{amssymb}
\usepackage{amsmath}
\usepackage{indentfirst}
\usepackage{epsfig}
\usepackage{feynmf}
\usepackage{epstopdf}
\usepackage{subfigure}
\usepackage{slashed}
\usepackage{cases}
\usepackage{multirow}
\usepackage[colorlinks,citecolor=blue,anchorcolor=red,menucolor=red,linkcolor=red,filecolor=red,runcolor=red,urlcolor=blue,frenchlinks=red]{hyperref}

\allowdisplaybreaks[3]

\begin{document}

\title{A hybrid nonet with $J^{PC}=1^{-+}$ or a tetraquark 81-plet}

\author{Niu Su$^1$}
\author{Er-Liang Cui$^2$}
\author{Yi-Wei Jiang$^3$}
\author{Hua-Xing Chen$^3$}
\email{hxchen@seu.edu.cn}

\affiliation{$^1$College of Physics and Optoelectronic Engineering, Taiyuan University of Technology, Taiyuan 030024, China
\\
$^2$College of Science, Northwest A\&F University, Yangling 712100, China
\\
$^3$School of Physics, Southeast University, Nanjing 210094, China}

\begin{abstract}
Confirming the existence of hybrid states remains challenging due to their experimental indistinguishability from tightly bound tetraquarks and loosely bound molecules. To address this issue, we employ QCD sum rules to systematically investigate the \( \pi_1(1600) \) and \( \eta_1(1855) \) as candidate tetraquark states with exotic quantum numbers \( J^{PC} = 1^{-+} \).  Within the hybrid framework, an \( SU(3) \) flavor nonet is expected, featuring two isoscalar configurations, \( q\bar{q}g \) and \( s\bar{s}g \), where \( q = u/d \). In contrast, the tetraquark scenario predicts an \( SU(3) \) flavor 81-plet comprising three isoscalar states: \( qq\bar{q}\bar{q} \), \( qs\bar{q}\bar{s} \), and \( ss\bar{s}\bar{s} \). Our analysis yields a mass of \( 2.22^{+0.18}_{-0.26}~\mathrm{GeV} \) for the \( ss\bar{s}\bar{s} \) tetraquark state, which is expected to decay predominantly into the \( \phi\phi \) and \( \eta f_1(1420) \) final states. Therefore, experimental scrutiny of their invariant mass spectra is pivotal for distinguishing between hybrid and tetraquark interpretations.
\end{abstract}


\maketitle

\pagenumbering{arabic}

$\\$
{\it Introduction} --- A hadron is a composite subatomic particle composed of quarks and gluons, bound together by the strong interaction. Hadrons are generally categorized into two families: mesons and baryons. In the traditional quark model, a meson is described as a bound state of a quark and an antiquark, while a baryon consists of three quarks. Despite its simplicity, this model has achieved remarkable success in describing the observed properties of hadrons and providing a coherent classification scheme~\cite{pdg}. Hadrons that cannot be well described within the traditional quark model are referred to as \textit{exotic hadrons}. Among these, states with exotic quantum numbers \( J^{PC} = 1^{-+} \) are of particular interest, as such quantum numbers are forbidden for conventional \( \bar{q}q \) mesons~\cite{Amsler:2004ps,Klempt:2007cp,Meyer:2015eta,Chen:2022asf}. Moreover, several experimental studies have reported evidence for the existence of such states:
\begin{itemize}

\item Three isovector states with exotic quantum numbers \( I^GJ^{PC} = 1^-1^{-+} \) have been reported: \( \pi_1(1400) \)~\cite{IHEP-Brussels-LosAlamos-AnnecyLAPP:1988iqi}, \( \pi_1(1600) \)~\cite{E852:1998mbq}, and \( \pi_1(2015) \)~\cite{E852:2004gpn}. Among these, the \( \pi_1(2015) \) was only observed in the BNL E852 experiment~\cite{E852:2004gpn}, and its existence remains inconclusive; we therefore do not include it in the present study. Furthermore, the COMPASS and JPAC collaborations have analyzed the \( \eta^{(\prime)}\pi \) decay channels~\cite{JPAC:2018zyd,COMPASS:2021ogp}, and their results suggest the presence of only one exotic \( \pi_1 \) resonance that couples to both channels, with no evidence for a second state. This implies that the \( \pi_1(1400) \) and \( \pi_1(1600) \) may correspond to the same underlying resonance. The mass and width of this state, identified as the \( \pi_1(1600) \), were determined to~\cite{JPAC:2018zyd}:
\begin{eqnarray}
\pi_1(1600) &:& M = 1564 \pm 24 \pm 86~\mathrm{MeV}/c^2 \, , \\
\nonumber && \Gamma = 492 \pm 54 \pm 102~\mathrm{MeV} \, .
\end{eqnarray}

\item The BESIII collaboration performed a partial wave analysis of the \( J/\psi \to \gamma \eta \eta^\prime \) decay and observed the \( \eta_1(1855) \) in the \( \eta \eta^\prime \) invariant mass spectrum, with a statistical significance exceeding \( 19\sigma \)~\cite{BESIII:2022riz}. This state possesses the exotic quantum numbers \( I^GJ^{PC} = 0^+1^{-+} \), and its mass and width were measured to be
\begin{eqnarray}
\eta_1(1855) &:& M = 1855 \pm 9 ^{+6}_{-1}~\mathrm{MeV}/c^2 \, , \\
\nonumber && \Gamma = 188 \pm 18 ^{+3}_{-8}~\mathrm{MeV} \, .
\end{eqnarray}

\end{itemize}

\begin{figure}[]
\begin{center}
\includegraphics[width=0.49\textwidth]{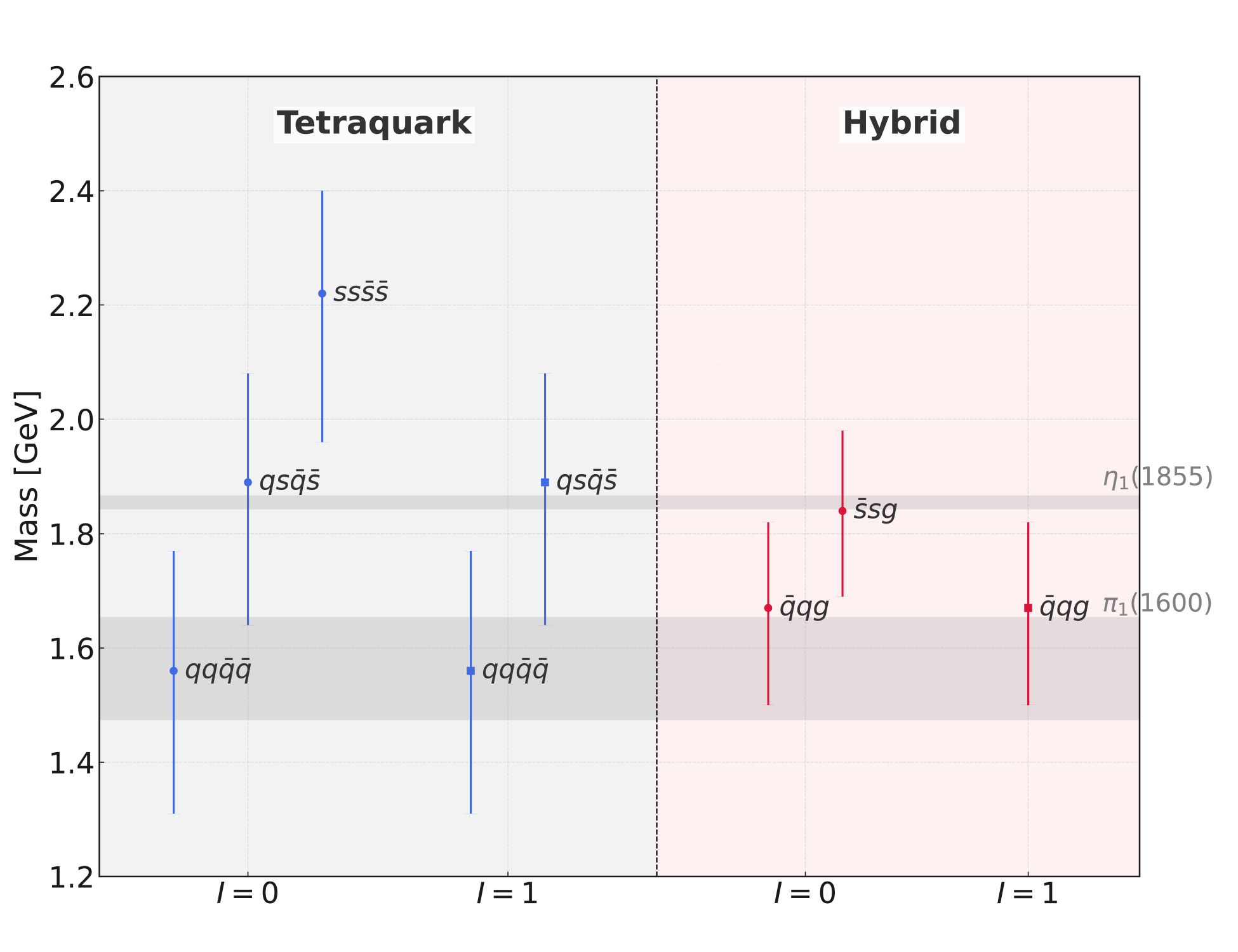}
\caption{Comparison of the tetraquark and hybrid spectra with exotic quantum numbers \( J^{PC} = 1^{-+} \) ($q=u/d$), as calculated from QCD sum rules in this study and in Ref.~\cite{Tan:2024grd}.}
\label{fig:phase}
\end{center}
\end{figure}

In this Letter we focus on the \( \pi_1(1600) \) and \( \eta_1(1855) \), whose possible internal structures have been widely discussed in the literature. Proposed configurations include hybrid states~\cite{Chen:2022isv,Qiu:2022ktc,Shastry:2022mhk,Chen:2023ukh,Esmer:2025xss}, tightly bound tetraquarks~\cite{Chen:2008qw,Chen:2008ne,Wan:2022xkx}, loosely bound molecules~\cite{Dong:2022cuw,Yang:2022rck,Liu:2024lph}, and other exotic configurations. Among these, the hybrid interpretation has attracted particular attention over the past five decades~\cite{Chanowitz:1982qj,Frere:1988ac,Page:1998gz,Iddir:2007dq,Wang:2025ypo}. In particular, we have applied the QCD sum rule method to systematically study these states in Refs.~\cite{Chen:2010ic,Chen:2022qpd,Tan:2024grd}. As illustrated in Fig.~\ref{fig:phase}, the hybrid framework predicts an \( SU(3) \) flavor nonet containing two isoscalar states, \( | q \bar{q} g; 0^+1^{-+} \rangle \) and \( | s \bar{s} g; 0^+1^{-+} \rangle \) (with \( q = u/d \)), as well as one isovector state, \( | q \bar{q} g; 1^-1^{-+} \rangle \). Within this framework, the \( \pi_1(1600) \) can be understood as the isovector state \( | q \bar{q} g; 1^-1^{-+} \rangle \), while the \( \eta_1(1855) \) may correspond to either of the isoscalar states.

However, distinguishing hybrid states from tightly bound tetraquarks or loosely bound molecules remains a major experimental and theoretical challenge. Due to the relatively large width of the \( \pi_1(1600) \), its molecular interpretation is disfavored, and in this work we concentrate on the tetraquark scenario. We perform a systematic analysis of tetraquark states with exotic quantum numbers \( J^{PC} = 1^{-+} \) using the QCD sum rule approach. As depicted in Fig.~\ref{fig:phase}, the tetraquark framework predicts an \( SU(3) \) flavor 81-plet. This multiplet includes three isoscalar states, \( | qq\bar{q}\bar{q}; 0^+1^{-+} \rangle \), \( | qs\bar{q}\bar{s}; 0^+1^{-+} \rangle \), and \( | ss\bar{s}\bar{s}; 0^+1^{-+} \rangle \), as well as two isovector states, \( | qq\bar{q}\bar{q}; 1^-1^{-+} \rangle \) and \( | qs\bar{q}\bar{s}; 1^-1^{-+} \rangle \). 

In the tetraquark framework, the \( \pi_1(1600) \) is interpreted as the isovector state \( | qq\bar{q}\bar{q}; 1^-1^{-+} \rangle \), and the \( \eta_1(1855) \) corresponds to the isoscalar state \( | qs\bar{q}\bar{s}; 0^+1^{-+} \rangle \). Furthermore, this framework predicts the existence of a purely strange isoscalar state, \( | ss\bar{s}\bar{s}; 0^+1^{-+} \rangle \), with an estimated mass of \( 2.22^{+0.18}_{-0.26}~\mathrm{GeV} \). Since both this state and the \( \eta_1(1855) \) are interpreted as tetraquark states, their decay widths are expected to be comparable, making the former a promising candidate for experimental detection. This state is anticipated to decay predominantly into the \( \phi\phi \) and \( \eta f_1(1420) \) final states. Therefore, searches for resonances in the corresponding invariant mass spectra would be essential not only for validating the tetraquark interpretation, but also for probing the potential existence of hybrid configurations.

$\\$
{\it QCD sum rule analysis} --- To begin with, we construct the $ss\bar{s}\bar{s}$ tetraquark currents with exotic quantum numbers $J^{PC} = 1^{-+}$. There exist two independent diquark–antidiquark currents:
\begin{eqnarray}
\eta_{1\mu} &=& (s_a^T C \gamma_5 s_b)(\bar{s}_a \gamma_\mu \gamma_5 C \bar{s}_b^T)
\label{def:eta1} 
\\ \nonumber &+& (s_a^T C \gamma_\mu \gamma_5 s_b)(\bar{s}_a \gamma_5 C \bar{s}_b^T) \, ,
\\
\eta_{2\mu} &=& (s_a^T C \gamma^\nu s_b)(\bar{s}_a \sigma_{\mu\nu} C \bar{s}_b^T) 
\label{def:eta2}
\\ \nonumber &+& (s_a^T C \sigma_{\mu\nu} s_b)(\bar{s}_a \gamma^\nu C \bar{s}_b^T)\, ,
\end{eqnarray}
where \( a \) and \( b \) are color indices, \( C = i \gamma_2 \gamma_0 \) is the charge-conjugation matrix, and repeated Lorentz indices are implicitly summed.

We apply the QCD sum rule method~\cite{Shifman:1978bx,Reinders:1984sr} to analyze these currents by evaluating the two-point correlation function
\begin{eqnarray}   
\Pi^{ij}_{\mu\nu}(q) &\equiv& i \int d^4x\, e^{iq\cdot x} \langle 0 | \mathcal{T}[ \eta_{i\mu}(x) \eta^\dagger_{j\nu}(0)] | 0 \rangle 
\label{def:pi}
\\ \nonumber
&=& \left(g_{\mu\nu} - \frac{q_\mu q_\nu}{q^2} \right) \times \Pi_{ij}(q^2) + \cdots \, ,
\end{eqnarray}
where we have projected out the Lorentz structure relevant to a vector state. We compute this correlation function at the quark–gluon level using the method of operator product expansion (OPE), and find that the off-diagonal components vanish:
\begin{equation}   
\Pi^{12}_{\mu\nu}(q) = \Pi^{21}_{\mu\nu}(q) = 0 \, .
\end{equation}
As discussed in Ref.~\cite{Chen:2018kuu}, this result indicates that the two currents \( \eta_{1\mu} \) and \( \eta_{2\mu} \) are orthogonal and therefore do not significantly overlap with the same physical state. We thus assume that they couple to two distinct states, denoted \( X_1 \) and \( X_2 \), respectively.

We take the first current $\eta_{1\mu}$ as an example to calculate the mass \( M_1 \) of the associated state \( X_1 \). The correlation function $\Pi_{11}(q^2)$ can be expressed via a dispersion relation
\begin{equation}
\Pi_{11}(q^2) = \int_{s_<}^\infty \frac{\rho_{11}(s)}{s - q^2 - i\varepsilon}\, ds \, ,
\end{equation}
where $s_<=16 m_s^2$ is the physical threshold, and \( \rho_{11}(s) \equiv \operatorname{Im} \Pi_{11}(s) / \pi \) denotes the corresponding spectral density.

At the hadron level, the spectral density is obtained by inserting a complete set of intermediate states into the correlation function:
\begin{eqnarray}
&& \rho^{\rm phen}_{11}(s) \times \left(g_{\mu\nu} - \frac{q_\mu q_\nu}{q^2}\right)
\label{eq:rho}
\\ \nonumber 
&=& \sum_n \delta(s - M_n^2) \langle 0 | \eta_{1\mu} | n \rangle \langle n | \eta_{1\nu}^\dagger | 0 \rangle
\\ \nonumber 
&=& f_1^2 \delta(s - M_1^2) \times \left(g_{\mu\nu} - \frac{q_\mu q_\nu}{q^2}\right) + \cdots \, ,
\end{eqnarray}
where we adopt the single-pole dominance approximation for the lowest-lying resonance \( X_1 \). The coupling constant \( f_1 \) is defined via the matrix element
\begin{equation}
\langle 0 | \eta_{1\mu} | X_1 \rangle = f_1 \, \epsilon_\mu \, ,
\label{coupling1}
\end{equation}
with \( \epsilon_\mu \) being the polarization vector of the vector meson.

At the quark–gluon level, the spectral density is evaluated using the method of operator product expansion (OPE). The calculation is carried out up to dimension-12 ($D=12$) terms and includes contributions from the perturbative term, the quark condensate, the gluon condensate, the quark–gluon mixed condensate, and their combinations. The resulting expression reads:
\begin{eqnarray}
\rho_{11}^{\rm OPE}(s) &=& {s^4 \over 18432 \pi^6}
- {19 m_s^2\over3840\pi^6} s^3 
\\ \nonumber
&-& \Big ( {\langle g_s^2 GG \rangle\over18432\pi^6 }
- {m_s\langle \bar s s \rangle\over16\pi^4}\Big )s^2-\Big({\langle \bar s s \rangle^2\over6\pi^2}
\\ \nonumber
&-& {17m_s^2\langle g_s^2 GG \rangle\over9216\pi^6 }
- {7m_s\langle g_s \bar s \sigma G s \rangle\over96\pi^4}
\Big)s
\\ \nonumber
&-&{m_s \langle g_s^2 GG \rangle \langle \bar s s \rangle \over 128 \pi^4 } 
+{ m_s^2\langle \bar s s \rangle^2 \over 4 \pi^2 }
\\ \nonumber
&-&{7 \langle \bar s s \rangle \langle g_s \bar s \sigma G s \rangle \over 24 \pi^2 }
+\Big( 
- {5 m_s^2 \langle \bar s s \rangle \langle g_s \bar s \sigma G s \rangle \over 8\pi^2}
\\ \nonumber
&-&{5 \langle g_s^2 GG \rangle  \langle \bar s s \rangle^2 \over 864 \pi^2}
+{5 m_s \langle g_s^2 GG \rangle  \langle g_s \bar s \sigma G s \rangle \over 2304 \pi^4} 
\\ \nonumber
&+& { \langle g_s \bar s \sigma G s \rangle^2 \over 12\pi^2 } -{4 m_s \langle \bar s s \rangle^3   \over 9 }\Big)\delta(s-s_<)
\\ \nonumber
&-&\Big({13 m_s^2 \langle g_s \bar s \sigma G s \rangle^2 \over 48\pi^2}+{m_s^2 \langle g_s^2 GG \rangle  \langle \bar s s \rangle^2 \over 576 \pi^2}
\\ \nonumber
&-&{ \langle g_s^2 GG \rangle \langle \bar s s \rangle \langle g_s \bar s \sigma G s \rangle \over 576 \pi^2} 
\\ \nonumber 
&-& {37 m_s \langle \bar s s \rangle^2 \langle g_s \bar s \sigma G s \rangle  \over 18 }\Big)\delta^\prime(s-s_<).
\end{eqnarray}

By equating the spectral densities at the hadron and quark–gluon levels, and subsequently applying a Borel transformation to suppress contributions from higher excited states and the continuum, we arrive at the QCD sum rule equation
\begin{eqnarray}
\Pi_{11}(s_0, M_B^2) &=& f_1^2 e^{-M_1^2 / M_B^2}
\\ \nonumber
&=& \int_{s_<}^{s_0} e^{-s / M_B^2} \rho_{11}^{\rm OPE}(s) \, ds \, ,
\label{eq_fin}
\end{eqnarray}
where the continuum contribution is modeled by the OPE spectral density above the threshold value \( s_0 \), based on the quark–hadron duality assumption. The mass \( M_1 \) of the state $X_1$ is be extracted as
\begin{equation}
M_1^2(s_0, M_B^2) = 
\frac{ \int_{s_<}^{s_0} e^{-s / M_B^2} \, s \, \rho_{11}^{\rm OPE}(s) \, ds }
     { \int_{s_<}^{s_0} e^{-s / M_B^2} \, \rho_{11}^{\rm OPE}(s) \, ds } \, .
\label{eq:LSR}
\end{equation}

$\\$
{\it Numerical analysis} --- We continue with the current \(\eta_{1\mu}\) as an example to perform the numerical analysis. The QCD input parameters adopted in the calculation are summarized as follows~\cite{pdg,Yang:1993bp,Gimenez:2005nt,Narison:2018dcr}:
\begin{eqnarray}
\nonumber m_s(2\,\mathrm{GeV}) &=& 93.5\pm0.5 \, \mathrm{MeV} \,,
\\ \langle \bar{s} s \rangle &=& -(0.8 \pm 0.1) \times (0.240\,\mathrm{GeV})^3 \,,
\label{condensates}
\\ \nonumber \langle g_s \bar{s} \sigma G s \rangle &=& - 0.8 \times \langle \bar{s} s \rangle \,,
\\ \nonumber \langle \alpha_s G G \rangle &=& (6.35 \pm 0.35) \times 10^{-2} \, \mathrm{GeV}^4 \,.
\end{eqnarray}
To reliably extract the mass using Eq.~(\ref{eq:LSR}), it is essential to determine appropriate working regions for the two free parameters: the threshold value \( s_0 \) and the Borel mass \( M_B \).

First, we ensure good convergence of OPE by requiring that the contribution from dimension-12 condensates be less than 5\% of the total, and that from dimension-10 condensates be less than 10\%:
\begin{eqnarray}
\mathrm{CVG} &\equiv& \left| \frac{\Pi_{11}^{D=12}(\infty, M_B^2)}{\Pi_{11}(\infty, M_B^2)} \right| \leq 5\% \,,
\\
\mathrm{CVG}^\prime &\equiv& \left| \frac{\Pi_{11}^{D=10}(\infty, M_B^2)}{\Pi_{11}(\infty, M_B^2)} \right| \leq 10\% \,.
\label{eq:convergence}
\end{eqnarray}
This condition imposes a lower bound on the Borel mass: $M_B^2 \geq 1.56$~GeV$^2$.

Second, to ensure dominance of the ground-state pole contribution, we require that the pole contribution (PC) be larger than 40\%:
\begin{equation}
\mathrm{PC} \equiv \left| \frac{\Pi_{11}(s_0, M_B^2)}{\Pi_{11}(\infty, M_B^2)} \right| \geq 40\% \,.
\label{eq:pole}
\end{equation}
We find that this condition is satisfied only when $s_0 \geq 7.7$~GeV$^2$. Accordingly, we adopt a slightly larger value \( s_0 = 8.5\, \mathrm{GeV}^2 \), which results in a valid Borel window of $1.56$~GeV$^2 \leq M_B^2 \leq 1.72$~GeV$^2$.

\begin{figure*}[]
\begin{center}
\subfigure[]{\includegraphics[width=0.4\textwidth]{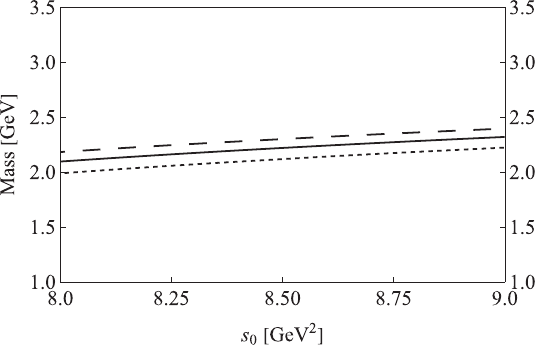}}
~~~~~~~~~~
\subfigure[]{\includegraphics[width=0.4\textwidth]{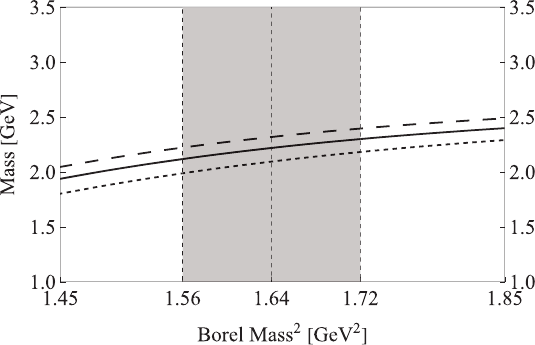}}
\caption{Dependence of the mass \( M_1 \) on (a) the threshold value \( s_0 \) and (b) the Borel mass \( M_B \). In the left panel, the short-dashed, solid, and long-dashed curves correspond to \( M_B^2 = 1.56,\,1.64,\,1.72~\mathrm{GeV}^2 \), respectively. In the right panel, the curves correspond to \( s_0 = 8.0,\,8.5,\,9.0~\mathrm{GeV}^2 \), respectively.}
\label{fig:eta1mass}
\end{center}
\end{figure*}

Finally, we require that the extracted mass \( M_1 \) exhibit only mild sensitivity to variations in both the threshold value \( s_0 \) and the Borel mass \( M_B \). As shown in Fig.~\ref{fig:eta1mass}, the mass dependence on these parameters remains moderate within the working regions $8.0$~GeV$^2 \leq s_0 \leq 9.0$~GeV$^2$ and $1.56$~GeV$^2 \leq M_B^2 \leq 1.72$~GeV$^2$. The resulting mass is determined to be
\begin{equation}
M_1 = 2.22^{+0.18}_{-0.26}~\mathrm{GeV} \,,
\end{equation}
where the uncertainty reflects the combined variations in the Borel mass \( M_B \), the threshold value \( s_0 \), and the QCD input parameters listed in Eqs.~(\ref{condensates}).

Similarly, we perform a QCD sum rule analysis using the current \(\eta_{2\mu}\) and extract the mass \( M_2 \) of the associated state \( X_2 \):
\begin{equation}
M_2 = 2.62^{+0.09}_{-0.11}~\mathrm{GeV} \,.
\end{equation}
This value is significantly larger than \( M_1 \), indicating a notable difference between the two currents. Given this disparity, we do not include the result obtained from \(\eta_{2\mu}\) in the subsequent discussions.

By appropriately modifying the quark content, our analysis can be extended to extract the masses of other tetraquark states with exotic quantum numbers \( J^{PC} = 1^{-+} \). As summarized in Fig.~\ref{fig:phase}, three isoscalar and two isovector states emerge, with their masses calculated as follows (\( q = u/d \)):
\begin{eqnarray}
\nonumber | q q \bar{q} \bar{q}; 0^+ 1^{-+} / 1^- 1^{-+} \rangle &:& M = 1.56^{+0.21}_{-0.25}~\mathrm{GeV} \,,
\\ | q s \bar{q} \bar{s}; 0^+ 1^{-+} / 1^- 1^{-+} \rangle &:& M = 1.89^{+0.19}_{-0.25}~\mathrm{GeV} \,,
\\ \nonumber | s s \bar{s} \bar{s}; 0^+ 1^{-+} \rangle &:& M = 2.22^{+0.18}_{-0.26}~\mathrm{GeV} \,.
\end{eqnarray}
Accordingly, our results suggest that the \(\pi_1(1600)\) can be interpreted as the isovector tetraquark state \(| q q \bar{q} \bar{q}; 1^- 1^{-+} \rangle\), while the recently observed \(\eta_1(1855)\) may be described as the isoscalar tetraquark state \(| q s \bar{q} \bar{s}; 0^+ 1^{-+} \rangle\).

$\\$
{\it Decay properties} --- Besides the diquark–antidiquark currents \(\eta_{1\mu}\) and \(\eta_{2\mu}\), defined in Eqs.~(\ref{def:eta1}) and (\ref{def:eta2}), there also exist two independent mesonic–mesonic currents:
\begin{eqnarray}
\eta_{3\mu} &=& (\bar{s}_a \gamma_5 s_a)(\bar{s}_b \gamma_\mu \gamma_5 s_b) \,,
\\ \eta_{4\mu} &=& (\bar{s}_a \gamma^\nu s_a)(\bar{s}_b \sigma_{\mu\nu} s_b) \,.
\end{eqnarray}
These currents are related to the diquark–antidiquark currents via the Fierz rearrangement:
\begin{eqnarray}
\eta_{1\mu} &=& -\eta_{3\mu} + i\, \eta_{4\mu} \,,
\\ \nonumber
\eta_{2\mu} &=& -3i\, \eta_{3\mu} + \eta_{4\mu} \,.
\label{eq:fierz2}
\end{eqnarray}
This relation demonstrates an equivalence between diquark–antidiquark and mesonic–mesonic configurations at the level of local interpolating operators, enabling one to choose between different current representations when studying decay properties~\cite{Chen:2021erj}.

\begin{figure}[]
\begin{center}
\includegraphics[width=0.25\textwidth]{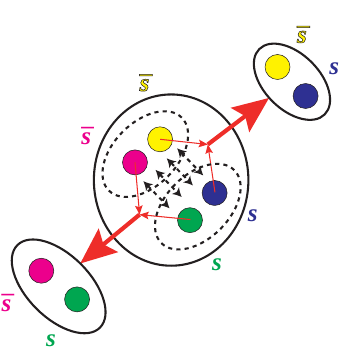}
\caption{Decay process of the tetraquark state $|ss\bar{s}\bar{s};\,0^+1^{-+}\rangle$, coupled by the current $\eta_{1\mu}$, into two strangeonium mesons.}
\label{fig:decay}
\end{center}
\end{figure}

As illustrated in Fig.~\ref{fig:decay}, We again take the current \(\eta_{1\mu}\) as an example and study its coupling to the \(\phi \phi\) final state. The relevant matrix element reads
\begin{eqnarray}
\langle 0 | \eta_{1\mu} | \phi \phi \rangle
&=& i\, \langle 0 | \eta_{4\mu} | \phi \phi \rangle
\\ \nonumber
&=& 2i\, \langle 0 | \bar{s}_a \gamma^\nu s_a | \phi \rangle \langle 0 | \bar{s}_b \sigma_{\mu\nu} s_b | \phi \rangle \,.
\end{eqnarray}
Here, the factor of 2 arises from the symmetry under exchange of the two identical \(\phi\) mesons. It is important to note that this calculation is performed within the naive factorization framework, which is known to introduce substantial theoretical uncertainties compared to more sophisticated methods such as QCD factorization~\cite{Beneke:2001ev}. Nonetheless, many of these uncertainties tend to cancel when considering relative branching ratios, making this approach useful for qualitative estimates.

The strangeonium operators couple to the \(\phi\) meson through the following matrix elements~\cite{Becirevic:2003pn,RBC-UKQCD:2008mhs}:
\begin{eqnarray}
\langle 0 |\bar{s}_a \gamma_\mu s_a|\phi(p,\epsilon)\rangle
&=& m_{\phi} f_{\phi} \epsilon_{\mu} \,,
\\ \nonumber
\langle 0 |\bar{s}_a \sigma_{\mu\nu} s_a|\phi(p,\epsilon)\rangle
&=& i f_{\phi}^T (p_{\mu}\epsilon_{\nu} - p_{\nu}\epsilon_{\mu}) \,.
\end{eqnarray}
Using these relations, we can derive the decay amplitude for the process in which the state $X_1$, coupled by the current $\eta_{1\mu}$, decays into two $\phi$ mesons:
\begin{eqnarray}
&& \langle X_1(p,\epsilon) | \phi(p_1,\epsilon_1)\phi(p_2,\epsilon_2) \rangle
\\ \nonumber &\propto& - 2 m_{\phi} f_{\phi} f_{\phi}^T \epsilon^\mu \epsilon_1^\nu (p_{2\mu} \epsilon_{2\nu} - p_{2\nu} \epsilon_{2\mu}) \, .
\end{eqnarray}
From this amplitude, the partial decay width \(\Gamma_{X_1 \to \phi \phi}\) can be calculated.

In a similar manner, we evaluate the partial decay widths for other final states, including \(\eta \eta^\prime\) and \(\eta f_1(1420)\). By comparing these results, we obtain the following relative branching fractions:
\begin{eqnarray}
\nonumber \mathcal{B}(\,|ss\bar{s}\bar{s};\,0^+1^{-+}\rangle &\to& ~\phi\phi~ :~ \eta\eta^\prime~ : ~\eta f_1(1420) \,)
\\ &=& ~~1~ :\, 0.004 \,: ~~~ 1.2 \, .
\end{eqnarray}
This indicates that the isoscalar tetraquark state \(|ss \bar{s} \bar{s};\,0^+1^{-+}\rangle\) decays predominantly into the \(\phi \phi\) and \( \eta f_1(1420) \) channels. 

The same methodology can be extended to other tetraquark states. A detailed analysis of their decay properties will be presented in future studies.

$\\$
{\it Summary and Discussions} --- In this Letter we investigate the \( \pi_1(1600) \) and \( \eta_1(1855) \), both characterized by the exotic quantum numbers \( J^{PC} = 1^{-+} \), which are forbidden for conventional quark--antiquark mesons. These states are considered prime candidates for nonstandard configurations, including hybrid mesons, tightly bound tetraquarks, or loosely bound molecules. While the hybrid interpretation has been extensively explored in the literature, the tetraquark scenario remains particularly compelling due to the rich internal dynamics inherent to four-quark systems.

Using QCD sum rules, we construct interpolating currents for tetraquark states with \( J^{PC} = 1^{-+} \). Our results support identifying the \( \pi_1(1600) \) as an isovector \( qq\bar{q}\bar{q} \) state and the \( \eta_1(1855) \) as an isoscalar \( qs\bar{q}\bar{s} \) state, where \( q = u/d \). In addition, we predict the existence of an isoscalar \( ss\bar{s}\bar{s} \) state with \( J^{PC} = 1^{-+} \) and a mass of \( 2.22^{+0.18}_{-0.26}~\mathrm{GeV} \). This state is expected to decay predominantly into the \( \phi\phi \) and \( \eta f_1(1420) \) channels. Moreover, since both the \( \eta_1(1855) \) and the predicted \( ss\bar{s}\bar{s} \) state are interpreted as tetraquarks, their total widths are expected to be comparable, implying that the latter may also be sufficiently narrow for experimental detection.

The observation of an exotic \( I^GJ^{PC} = 0^+1^{-+} \) state in the \( \phi\phi \) and \( \eta f_1(1420) \) channels would provide strong evidence in support of the tetraquark interpretation and help differentiate it from competing models. Conversely, the absence of such a signal would lend credence to the hybrid interpretation. Future high-statistics measurements at BESIII, PANDA, and GlueX will be essential for determining the masses, widths, and decay patterns of exotic states in the light and strange quark sectors. Such investigations will not only clarify the nature of these resonances but also test the existence of hybrid configurations, thereby advancing our understanding of nonperturbative QCD.

\section*{Acknowledgments}

This work is supported by
the National Natural Science Foundation of China under Grants No.~12005172 and No.~12075019,
the Jiangsu Provincial Double-Innovation Program under Grant No.~JSSCRC2021488,
and
the Fundamental Research Funds for the Central Universities.

\bibliographystyle{elsarticle-num}
\bibliography{ref}

\end{document}